\documentclass[superscriptaddress,nofootinbib, amsmath,amssymb, aps,prc,portrait]{revtex4-1}
\usepackage{multirow,dcolumn}
\usepackage{xcolor,colortbl}
\usepackage{longtable}
\usepackage{appendix}
\usepackage{graphicx}
\usepackage{bm}
\usepackage{hyperref}
\hypersetup{colorlinks=true, citecolor=blue, urlcolor=blue, linkcolor=blue}
\usepackage[normalem]{ulem}
\usepackage[T1]{fontenc}    

\newcommand{\mee}       {$m_{\beta\beta}$}

\newcommand{\BBz}       {$0\nu\beta\beta$}

\newcommand{\BBt}       {$2\nu\beta\beta$}
\newcommand{\BB}        {$\beta\beta$}
\newcommand{\Mz}        {$M_{0\nu}$}

\newcommand{\ttwo}      {$\theta_{12}$}

\newcommand{\qval}      {$Q_{\beta\beta}$}

\newcommand{\Tz}        {$T^{0\nu}_{1/2}$}

\newcommand{\cpFty}     {cts/(FWHM t yr)}

\newcommand{\nuc}[2]    {$^{#1}$\textrm{#2}} 

\newcommand{\LEG}       {LEGEND}
\newcommand{\Ltwo}      {{\LEG-200}}
\newcommand{\Lthou}     {{\LEG-1000}}

\newcommand{\MJ}        {\textsc{Majorana}}
\newcommand{\DEM}       {\textsc{Demonstrator}}

\newcommand{\Gerda}     {\textsc{Gerda}}

\newcommand{\be}        {\begin{equation}}
\newcommand{\ee}        {\end{equation}}

\definecolor{Gray}{gray}{0.85}
\definecolor{LightCyan}{rgb}{0.88,1,1}
\definecolor{BlueTable}{rgb}{0.30,0.58,0.93}
\definecolor{SREblizzardblue}{rgb}{0.9, 0.9, 0.98}
\definecolor{powderblue}{rgb}{0.69, 0.88, 0.9}
\definecolor{turquoiseblue}{rgb}{0.0, 1.0, 0.94}
\definecolor{skyblue}{rgb}{0.53, 0.81, 0.92}
\definecolor{lightskyblue}{rgb}{0.53, 0.81, 0.98}
\definecolor{lightcornflowerblue}{rgb}{0.6, 0.81, 0.93}
\definecolor{SREblizzardbluemist}{rgb}{0.9, 0.9, 0.98}
\definecolor{SREblizzardblue}{rgb}{0.0, 1.0, 1.0}
\definecolor{SREblizzardblue}{rgb}{0.91, 1.0, 1.0}
\definecolor{SREblizzardblue}{rgb}{.8, 0.9, 0.93}

\newcommand{\nucl}[2]{\ensuremath {}^{#2}\mathrm{#1}}

\begin{document}

\newcommand{\McGill}{McGill University, Department of Physics, Montreal, QC H3A 2T8, Canada}
\newcommand{\TRIUMF}{TRIUMF, Vancouver, BC V6T 2A3, Canada}
\newcommand{\NCSU}{North Carolina State University, Raleigh, NC 27695, USA}
\newcommand{\LANL}{Los Alamos National Laboratory, Los Alamos, NM 87545, USA}
\newcommand{\LLNL}{Lawrence Livermore National Laboratory, Livermore, CA 94550, USA}
\newcommand{\Duke}{Duke University, Department of Physics, Durham, NC 27708, USA}
\newcommand{\Yale}{Wright Laboratory, Department of Physics, Yale University, New Haven, CT 06520, USA}
\newcommand{\UTArlington}{Department of Physics, University of Texas at Arlington, Arlington, TX 76019, USA}
\newcommand{\MSU}{Facility for Rare Isotope Beams and Department of Physics and Astronomy, Michigan State University, East Lansing, MI 48824-1321, USA}
\newcommand{\ORNL}{Oak Ridge National Laboratory, Oak Ridge, TN 37830, USA}
\newcommand{\UNC}{Department of Physics and Astronomy, University of North Carolina, Chapel Hill, NC 27599, USA}
\newcommand{\TUNL}{Triangle Universities Nuclear Laboratory, Durham, NC 27708, USA}
\newcommand{\CENPAUW}{Center for Experimental Nuclear Physics and Astrophysics and Department of Physics, University of Washington, Seattle, WA 98115, USA}
\newcommand{\INT}{Institute for Nuclear Theory, University of Washington, Seattle WA 98195, USA}
\newcommand{\Penn}{University of Pennsylvania, Philadelphia, PA 19104, USA}
\newcommand{\UCLA}{University of California, Los Angeles, CA 90095-1547, USA}
\newcommand{\UCSD}{Physics Department, University of California San Diego, La Jolla, CA 92093, USA}
\newcommand{\UniPd}{Dipartimento di Fisica e Astronomia dell'Universit\'{a} di Padova, Italy}
\newcommand{\INFNPd}{INFN Padova, Padova, Italy}
\newcommand{\Stanford}{Stanford University, Stanford, CA 94305, USA}
\newcommand{\IU}{Center for Exploration of Energy and Matter and Department of Physics, Indiana University, Bloomington, IN 47405, USA}
\newcommand{\SLAC}{SLAC National Accelerator Laboratory, Menlo Park, CA 94025, USA}
\newcommand{\LBNL}{Lawrence Berkeley National Laboratory, Berkeley, CA 94720, USA}
\newcommand{\NRCKI}{National Research Center ``Kurchatov Institute'', 117218 Moscow, Russia}
\newcommand{\USD}{University of South Dakota, Vermillion, SD 57069, USA}
\newcommand{\SDM}{South Dakota Mines, Rapid City, SD 57701, USA}
\newcommand{\Bratislava}{Department of Nuclear Physics and Biophysics, Comenius University, Bratislava, Slovakia}
\newcommand{\RomoTre}{Roma Tre University and INFN, Sez. di Roma Tre, I-00146 Rome, Italy}
\newcommand{\UCB}{Department of Physics, University of California, Berkeley, CA 94720-7300, USA}
\newcommand{\TNTech}{Tennessee Tech University, Cookeville, TN 38505, USA}
\newcommand{\UniMi}{Department of Physics, University of Milano, Milano, Italy}
\newcommand{\INFNMi}{INFN Milano, Milano, Italy}
\newcommand{\IEAP}{Institute of Experimental and Applied Physics, Czech Technical University in Prague, CZ-11000 Prague, Czech Republic}
\newcommand{\UTexas}{University of Texas at Austin, Austin, TX 78712, USA}
\newcommand{\UNM}{University of New Mexico, Albuquerque, NM 87131, USA}
\newcommand{\IKZ}{Leibniz-Institut f\"{u}r Kristallz\"{u}chtung, 12489 Berlin, Germany}
\newcommand{\nikhef}{Nikhef and the University of Amsterdam, Science Park, 1098XG Amsterdam, Netherlands}
\newcommand{\UZH}{Physik-Institut, University of Zurich, Winterthurerstrasse 190, 8057 Switzerland}
\newcommand{\ANL}{Argonne National Laboratory, Lemont, IL 60439, USA}
\newcommand{\USC}{Department of Physics and Astronomy, University of South Carolina, Columbia, SC 29208, USA}
\newcommand{\MIT}{Massachusetts Institute of Technology, Cambridge, MA 02139, USA}
\newcommand{\CNRS}{Laboratoire SIMaP UMR 5266 CNRS-UGA-Grenoble INP, Saint Martin d'H\`{e}res, France}
\newcommand{\SAPIENZA}{Dipartimento di Fisica, Sapienza Universit\`{a} di Roma, Roma I-00185, Italy}
\newcommand{\INFNROMA}{INFN - Sezione di Roma, Roma I-00185, Italy}
\newcommand{\GSSI}{Gran Sasso Science Institute, L'Aquila, Italy}
\newcommand{\LNGS}{Istituto Nazionale di Fisica Nucleare, Laboratori Nazionali del Gran Sasso, Assergi (AQ), Italy}
\newcommand{\UNIVAQ}{Department of Physical and Chemical Sciences University of L'Aquila, L'Aquila, Italy}
\newcommand{\VTCNP}{Center for Neutrino Physics, Virginia Polytechnic Institute and State University, Blacksburg, Virginia 24061, USA}
\newcommand{\UniMiB}{Universit\`{a} di Milano Bicocca e sez. INFN di Milano Bicocca, Milano 20126, Italy}
\newcommand{\MiB}{INFN Sezione di Milano - Bicocca, Milano, Italy}
\newcommand{\CalPoly}{California Polytechnic State University, San Luis Obispo, CA 93407, USA}
\newcommand{\UA}{Department of Physics and Astronomy, University of Alabama, Tuscaloosa, AL 35487, USA}
\newcommand{\UMass}{University of Massachusetts, Amherst, MA 01003, USA} 
\newcommand{\FRIB}{Facility for Rare Isotope Beams, Michigan State University, East Lansing, MI 48824 USA}
\newcommand{\CoMines}{Colorado School of Mines, Golden, CO 80401 USA}
\newcommand{\RPI}{Department of Physics, Applied Physics and Astronomy, Rensselaer Polytechnic Institute, Troy, NY 12180, USA}
\newcommand{\PNNL}{Pacific Northwest National Laboratory, Richland, WA 99352, USA}
\newcommand{\Princeton}{Physics Department, Princeton University, Princeton, NJ 08544, USA}
\newcommand{\UWC} {Department of Physics and Astronomy, University of the Western Cape, P/B X17, Bellville 7535, South Africa}
\newcommand{\UBC}{University of British Columbia, Department of Physics and Astronomy, Vancouver, BC V6T 1Z1, Canada}
\newcommand{\UKY}{University of Kentucky, Lexington, KY, 40503}
\newcommand{\TUM}{Technical University of Munich, Munich, Germany}
\newcommand{\UTK}{University of Tennessee, TN 37996, USA}
\newcommand{\URegina}{Department of Physics, University of Regina, Regina, SK S4S 0A2, Canada}
\newcommand{\udel}{Department of Physics \& Astronomy, University of Delaware, Newark, DE 19716}
\newcommand{\tohokurcns}{Research Center for Neutrino Science, Tohoku University, Sendai, Miyagi 980-8578, Japan}
\newcommand{\bostonu}{Department of Physics, Boston University, Boston, MA 02215}
\newcommand{\Drexel}{Drexel University, Philadelphia, PA 19104, USA}
\newcommand{\Subatech}{SUBATECH, IMT Atlantique, CNRS/IN2P3, Nantes Universit\'{e}, Nantes 44307, France}
\newcommand{\IMECAS}{Institute of Microelectronics of the Chinese Academy of Sciences, Beijing, 100029, China}
\newcommand{\IHEP}{Institute of High Energy Physics, Beijing, 100049, China}
\newcommand{\JINR}{Joint Institute for Nuclear Research, Dubna, 141980, Russia}
\newcommand{\Jag}{M. Smoluchowski Institute of Physics, Jagiellonian University, 30-348 Krakow, Poland}
\newcommand{\MEPhI}{National Research Nuclear University MEPhI, Moscow, 115409, Russia}
\newcommand{\NU}{Northwestern University, Evanston, IL 60208, USA}
\newcommand{\SNOLAB}{SNOLAB, 1039 Regional Road 24, Lively, ON, P3Y 1N2, Canada}
\newcommand{\Carleton}{Physics Department, Carleton University, Ottawa, Ontario K1S 5B6, Canada}
\newcommand{\Iowa}{Department of Physics and Astronomy, Iowa State University, Physics Hall, 2323 Osborn Dr 12, Ames, IA 50011}
\newcommand{\FNAL}{Fermi National Accelerator Laboratory, Batavia, IL, USA}
\newcommand{\UNCW}{University of North Carolina Wilmington, Wilmington, NC 28403, USA}
\newcommand{\Laurentian}{Laurentian University, School of Natural Sciences, Sudbury ON P3 2C6, Canada}
\newcommand{\CSU}{Colorado State University, Fort Collins CO 80523, USA}
\newcommand{\UCL}{University College London, London, WC1E 6BT, UK}
\newcommand{\CEA}{IRFU, CEA, Universit\'e Paris-Saclay, 91191 Gif-sur-Yvette, France}
\newcommand{\UNIGE}{Dipartimento di Fisica, Universit\`a di Genova, I-16146, Italy}
\newcommand{\INFNGE}{INFN - Sezione di Genova, I-16146, Italy}
\newcommand{\IJCLab}{Universit\'e Paris-Saclay, CNRS/IN2P3, IJCLab, 91405 Orsay, France}
\newcommand{\SFU}{Department of Chemistry, Simon Fraser University, Burnaby, BC, Canada, V5A 1S6}
\newcommand{\jhu}{Johns Hopkins University, Baltimore, MD 21218, USA}
\newcommand{\Tuebingen}{University T\"{u}bingen, T\"{u}bingen, Germany}
\newcommand{\CMU}{Department of Physics, Central Michigan University, Mount Pleasant, MI 48859, USA}

\title{Neutrinoless Double Beta Decay}

\affiliation{\ANL}  
\affiliation{\VTCNP}
\affiliation{\SFU}
\affiliation{\Stanford}
\affiliation{\PNNL}
\affiliation{\UTArlington}
\affiliation{\USC}
\affiliation{\udel}
\affiliation{\NRCKI}
\affiliation{\Duke}
\affiliation{\TUNL} 
\affiliation{\UZH}
\affiliation{\SAPIENZA}
\affiliation{\INFNROMA}
\affiliation{\UCB}  
\affiliation{\UKY}  
\affiliation{\UniPd}
\affiliation{\INFNPd}
\affiliation{\MiB}
\affiliation{\TUM}
\affiliation{\SLAC}
\affiliation{\LLNL}
\affiliation{\UniMiB}
\affiliation{\RPI}
\affiliation{\McGill}
\affiliation{\TRIUMF}
\affiliation{\RomoTre}
\affiliation{\SNOLAB}
\affiliation{\IHEP}
\affiliation{\IMECAS}
\affiliation{\LBNL}
\affiliation{\Carleton} 
\affiliation{\UA}
\affiliation{\SDM}
\affiliation{\LANL}
\affiliation{\INT}
\affiliation{\UNCW}
\affiliation{\nikhef}
\affiliation{\ORNL}
\affiliation{\UNIVAQ}
\affiliation{\LNGS}
\affiliation{\CENPAUW}
\affiliation{\UNIGE}
\affiliation{\INFNGE}
\affiliation{\GSSI}
\affiliation{\Drexel}
\affiliation{\UTK}
\affiliation{\UNC}
\affiliation{\CSU}
\affiliation{\Laurentian}
\affiliation{\NU}
\affiliation{\UNM}
\affiliation{\MIT}
\affiliation{\UBC}
\affiliation{\IJCLab}
\affiliation{\bostonu}
\affiliation{\NCSU}
\affiliation{\URegina}
\affiliation{\CalPoly}
\affiliation{\Iowa}
\affiliation{\CMU}
\affiliation{\Yale}
\affiliation{\MSU}
\affiliation{\IEAP}
\affiliation{\UCLA}
\affiliation{\tohokurcns}
\affiliation{\Princeton}
\affiliation{\Tuebingen}
\affiliation{\Bratislava}
\affiliation{\TNTech}
\affiliation{\Penn}
\affiliation{\MEPhI}
\affiliation{\UMass}
\affiliation{\UTexas}
\affiliation{\FRIB}
\affiliation{\CoMines}
\affiliation{\UCSD}
\affiliation{\UWC}
\affiliation{\USD}
\affiliation{\Subatech}
\affiliation{\FNAL}
\affiliation{\CEA}
\affiliation{\IU}
\affiliation{\INFNMi}
\affiliation{\UniMi}
\affiliation{\UCL}
\affiliation{\jhu}
\affiliation{\IKZ}
\affiliation{\CNRS}
\affiliation{\JINR}
\affiliation{\Jag}
\affiliation{\UBC}

\author{C.~Adams}\affiliation{\ANL}
\author{K.~Alfonso}\affiliation{\VTCNP}
\author{C.~Andreoiu}\affiliation{\SFU}
\author{E.~Angelico}\affiliation{\Stanford}
\author{I.J.~Arnquist}\affiliation{\PNNL}
\author{J.A.A.~Asaadi}\affiliation{\UTArlington}
\author{F.T.~Avignone}\affiliation{\USC}
\author{S. N. Axani}\affiliation{\udel}
\author{A.S.~Barabash}\affiliation{\NRCKI}
\author{P.S.~Barbeau}\affiliation{\Duke}\affiliation{\TUNL}
% ...

\date{\today}

\begin{abstract} This White Paper, prepared for the Fundamental Symmetries, Neutrons, and Neutrinos Town Meeting related to the 2023 Nuclear Physics Long Range Plan, makes the case for double beta decay as a critical component of the future nuclear physics program. The major experimental collaborations and many theorists have endorsed this white paper.\footnote{These include CUPID, KamLAND-Zen, LEGEND, \textsc{Majorana}, nEXO, NEXT, SNO+, SuperNEMO, \textsc{Theia}.}
\end{abstract}

\maketitle

\section*{Executive Summary}

The discovery of neutrinoless double beta decay (\BBz) would reshape our fundamental understanding of neutrinos and of matter in the Universe. 
The search for \BBz\ tests whether there is a fundamental symmetry of Nature associated with Lepton Number, probes the quantum nature of neutrinos, and allows the measurement of their effective mass. It is the only practical way to demonstrate if neutrinos are their own antiparticles, that is, if neutrinos have a Majorana mass. 
The discovery of Majorana neutrinos would open the door to new physics beyond the discovery of neutrino oscillation, and would signify a paradigm shift in our understanding of the origins of mass and matter. The neutrino's non-zero mass impacts the evolution of the Universe from the beginning of time to the formation of large-scale structures in the present epoch, and Majorana neutrinos 
play a key role in 
viable scenarios that explain the matter-antimatter asymmetry in our Universe.

Several experimental approaches are now available to search with high sensitivity and low backgrounds for neutrinoless double beta decay in a variety of isotopes covering the entire region of the inverted mass ordering and beyond. Ton-scale experiments using large bolometer arrays (CUPID), high-resolution Ge detectors (LEGEND), and a large-volume liquid-Xe TPC (nEXO), have been identified as the leading next-generation experiments with US leadership. All three experiments are based on international collaborations leveraging the strengths of international partnerships and the world's premier underground laboratory facilities.  Those experiments are expected to extend the sensitivity to \BBz\ half lives by as much as two orders of magnitude.  During 2021, CUPID, LEGEND, and nEXO were examined in a double beta decay portfolio review organized by the DOE Office of Science Nuclear Physics
to address the opportunity for discovery of neutrinoless double beta decay,
covering the inverted ordering mass scale.
All three experiments were highly rated and judged to be worth pursuing.
R\&D challenges facing these three programs previously identified by a 2015 NSAC sub-committee have been resolved and CUPID, LEGEND, and nEXO are now preparing to proceed through the DOE Critical Decision process.  
The use of different isotopes and drastically different, yet robust and mature techniques is essential for validating a discovery of \BBz\ decay and mitigating any theoretical or experimental systematics.


If \BBz\ decay is discovered at the ton scale, advanced techniques will be required to probe the decay mechanism via topological information and event identification. If \BBz\ decay is not discovered, detectors that can reach greater exposures with improved background rejection will be required to attain sensitivity beyond the inverted mass ordering. A robust R\&D program is pursuing detector technologies with these capabilities. Among others, these include NEXT, which employs high pressure xenon gas time projection chambers with barium tagging, and \textsc{Theia}, a large-scale hybrid Cherenkov/scintillation detector that is an evolution of the techniques explored by the SNO+ and KamLAND-Zen experiments. 
With novel techniques and sensor technologies, rich reconstruction of event topologies, advanced particle identification, and half-life sensitivities in excess of $10^{28}$ years. The new detection capabilities of this future generation will also provide access to a wider physics program including probing CPT and baryon number violation, precision low-energy solar neutrino measurements, sensitivity to supernova neutrinos, and a rich array of other opportunities.

Research directed toward ongoing experiments (CUORE, KamLAND-Zen, LEGEND-200, SNO+, NEXT-100, SuperNEMO Demonstrator), which will have data in advance of the start of the ton-scale program, will push current limits for several isotopes, as well as acting as important test beds for techniques that can be used for experiments beyond the ton-scale.


In the 2015 Long Range Plan, a ton-scale experiment to search for neutrinoless double beta decay was considered the highest priority for new experiments in nuclear physics. Since then there has been great progress, including an order of magnitude improvement in the half-life sensitivity of existing experiments as well as advances on the various technologies yielding improved sensitivities. There has also been excellent progress in theory, towards the  understanding  nuclear matrix elements (NMEs) as well as possible new physics mechanisms, including quantification of the 
theoretical 
uncertainties. 
In addition, new detector technologies continue to be developed with the potential to probe neutrino masses beyond the inverse mass ordering. This R\&D is essential to maintain the leadership of US nuclear physics in \BBz\ science, and enable the next generation of discovery experiments. 

To enable a US-led program of discovery science, jointly with international partners, that could elucidate the nature of neutrinos and fundamentally alter our understanding of the origin of mass and matter in the Universe, we propose the following recommendation for the 2023 Long Range Plan in Nuclear Physics:

\noindent

\begin{itemize}

\item {\bf We recommend timely construction of ton-scale neutrinoless double beta decay experiments using multiple isotopes.}
\item {\bf We recommend support for a robust research program in neutrinoless double beta decay that includes the ongoing efforts in theory and experiment
as well as a diverse R\&D program exploring multiple promising isotopes and technologies with sensitivity beyond the inverted mass ordering.}
\end{itemize}

\newpage\newpage
\section{Scientific Motivation} \label{sec:introduction}

Neutrinoless double beta decay (\BBz) searches address,  {\it through a nuclear probe}, 
a number of  key open questions in subatomic physics and cosmology.   These include

\begin{itemize}
\item Is there a fundamental symmetry associated with Lepton Number? 
\item What is the origin of the neutrino's mass? 
Are neutrinos Majorana fermions, i.e.  their own antiparticles?
\item Why is there more matter than antimatter in the present universe?
\item What are the absolute masses of neutrinos, and how have they shaped the evolution of the universe?
\end{itemize}

In \BBz\ decay two neutrons convert into two protons with emission of two electrons and no neutrinos, thus changing the number of leptons by two units. Since lepton number $L$ (more precisely,  at the quantum level,  the difference $B-L$ of baryon and lepton number),   is conserved in the Standard Model,   observation of \BBz\ decay  would be direct evidence of new physics and 
would demonstrate that the neutrino mass has a Majorana component~\cite{Schechter:1981bd}, 
implying that neutrinos are self-conjugate, i.e. their own antiparticles. 
Observation of \BBz\ decay would also point to new mechanisms for mass generation, quite distinct from the one giving mass to other matter particles,  and possibly originating at very high energy scales. 
Finally,  the observation of a ``matter-creating'' process such as \BBz\ decay would 
 corroborate and  probe the so-called leptogenesis scenarios for the generation of the 
matter-antimatter asymmetry in the universe~\cite{Davidson:2008bu}. 





{Ton-scale  \BBz\ decay searches will probe at unprecedented levels a wide variety of mechanisms of lepton number violation. 
These mechanisms range from the high-scale seesaw naturally associated with Grand Unified Theories, to  models affecting electroweak scale physics, hence close to the TeV scale, all the way down to light (eV scale) right-handed neutrinos. 
This richness of physics enhances the discovery potential of \BBz\ decay. 
At the same time, this makes it hard  to quantify the discovery potential of the planned experiments in terms of a universal single metric.   
In this whitepaper we will use several quantities to  characterize the physics reach. First,
following standard practice we will focus on a class  of models 
in which \BBz\ is mediated by the exchange of the   three known light neutrinos, assuming they are  Majorana particles.  
In this case,  the decay rate  is proportional to  $G_{01} |M_{0\nu}|^2$$|$\mee $|^2$, 
where $G_{01}$ is a phase space factor, $M_{0\nu}$ is a nuclear matrix element and  \mee $= \sum_{i=1}^{3}  U_{ei}^2 m_i$ is the lepton number violating parameter, 
expressed in  terms of neutrino masses and elements $U_{ei}$ of the leptonic  mixing matrix. 
 \mee\ is partially determined by neutrino oscillation data~\cite{ParticleDataGroup:2022pth},  up to unknown CP-violating phases, 
 the overall neutrino mass scale,  and the normal or inverted ordering of the spectrum.
Therefore, in this scenario, very concrete discovery targets arise. 
Ton-scale experiments aim to cover the entire inverted ordering region, corresponding to \mee $> (18.4 \pm 1.3)$~meV,  and a discovery will be possible if the spectrum is inverted  or $m_{\rm lightest} > 50$~meV, irrespective of the ordering.  
Moreover, in this  scenario, falsifiable correlation arise with other neutrino mass probes, such as single beta decay and cosmology.

While it is common to present the physics reach of \BBz\ searches in terms of \mee\,  it is important to realize that this covers only one class 
of models for Majorana neutrino mass, in which lepton number violation (LNV) originates at very high mass scale $\Lambda$ 
and leaves behind $m_{\beta \beta} \sim v_{\rm ew}^2/\Lambda$  ($v_{\rm ew} \sim 200$~GeV is the Higgs expectation value) 
as  the only  low-energy footprint.
However, in many models  that incorporate Majorana neutrinos,  there exist other sources of LNV that 
can lead to sizable contributions to \BBz\ that are  {\it not directly related to the exchange of light neutrinos}. 
For example, in left-right symmetric models, apart from the exchange of light Majorana neutrinos, there appear contributions  from the exchange of heavy neutrinos,  heavy $W$ bosons,  and charged scalars with masses in the few TeV range.  
In full generality, at low-energy the effect of these heavy particles (in any model) is captured by a set of  $\Delta L=2$  local operators of odd dimension (seven, nine, ...), which are suppressed by odd powers of the heavy mass scale  $\Lambda$ associated with LNV ($1/\Lambda^3$,  $1/\Lambda^5$, ...). 
This is  analogous to the familiar Fermi theory of weak interactions, in which the effect of the $W$ exchange is captured at low energy  by the  usual $V-A$ current-current  (dimension six) interaction suppressed by $1/\Lambda_{\rm ew}^2$, with $\Lambda_{\rm ew} = 1/\sqrt{G_F}$. 
A systematic development of this effective field theory approach to LNV can be found in Ref.~\cite{Cirigliano:2018yza}.
In Fig.~\ref{fig:ExperimentComparision} we present the physics reach of current  and future \BBz\ searches in terms of 
both \mee\ and the scale $\Lambda$ associated to representative
dimension-seven and dimension-nine operators~\cite{Agostini2022MatterDiscover}, reaching well in the hundreds of TeV region, inaccessible to any other probe. 

In summary, given the breadth of mechanisms and scales associated with  lepton number violation,   
ton-scale searches for \BBz\  have a significant discovery potential that goes beyond the ``inverted mass ordering'' region in \mee, 
corresponding to} a plethora of models across the landscape of particle physics. 
As a result, other non-\BBz\ decay experimental efforts can complement results from \BBz\ decay searches, but do not diminish the need for further progress in \BBz\ decay science.


{\bf Mass Ordering Determined --}{If the mass ordering is determined, the inverted mass ordering could either be singled out or become irrelevant. Even in the latter case, the normal order branch still extends to high \mee\ values. Furthermore, lepton-number-violating processes other than light neutrino exchange are not constrained by oscillations. At present, even with the normal mass ordering, the probability of a discovery of \BBz\ decay is significant~\cite{Agostini2017}.}

{\bf Cosmological Probes of $\Sigma \equiv \sum_{i=1}^{3} m_i$ --}{Future efforts in observational cosmology aim to perform a first measurement of $\Sigma$. An observation of $\Sigma<100$~meV would effectively rule out the inverted mass ordering. Cosmology, however, does not discern the Majorana/Dirac character of the neutrino. A three-neutrino normal-ordering scenario with $\Sigma$ near its minimum would not constrain other potential lepton-number-violating processes that might contribute to
\BBz\ decay. Cosmology, as a standard model of physics with many parameters to be deduced, must be tested in all ways possible. There are few complementary laboratory measurements that can be done that directly test results from cosmology. Laboratory measurements of neutrino properties may provide such tests.}

{\bf Neutrino Mass Found --}{If neutrino mass is observed in $\beta$ decay, it will make the observation/non-observation of \BBz\ decay even more exciting. A null \BBz\ decay result might indicate Dirac neutrinos or, if lepton number violation is additionally observed in collider experiments, alternative / interfering mediation mechanisms or flavor symmetries driving $m_{\beta\beta}$ to be small.}

{\bf Lepton Number Violation Observed at Collider Experiments --}{The LHC or other collider experiments might observe lepton number violation consistent with a heavy neutrino, LR symmetry, or other BSM physics. Such a result would be complementary to a discovery of \BBz\ decay, leading to the interesting possibility of testing the underlying physics.}

{\bf Sterile Neutrinos --}{If a convincing demonstration of a sterile neutrino is found, it will fit well into the Majorana neutrino paradigm. This will increase \BBz\ decay interest. The new neutrino
might contribute to \BBz\ decay and significantly alter predicted \mee\ curves. (See for example Ref.~\cite{Barea2015}.) The 
sensitivity regions accessible to experiment will remain.}

{\bf Solar Mixing Angle --}{The interpretation of a limit or measurement of the rate of \BBz\ decay in terms of neutrino mass would be helped by better measurement of \ttwo.}

\subsection{A Scientific Opportunity with US Leadership} 

Many creative approaches to searches for \BBz\ decay have been undertaken, and three have reached the maturity and scale to be deployed as major projects in nuclear physics. 
The US scientists are playing leadership roles in many of the experimental and theoretical efforts. The next milestone in this quest is the development of experiments deploying isotopic mass on the ton scale, which will have the sensitivity to discover Majorana neutrinos with effective masses (\mee) as low as 10-20 meV, in the so-called inverted ordering region of neutrino masses.  Development of such an ambitious ton-scale program is both timely and scientifically relevant, as emphasized in the 2015 Long Range Plan~\cite{Geesaman-LRP-2015}.

Tremendous progress in pursuit of \BBz\ decay has been accomplished since the 2015 Long Range Plan. The limits on the \BBz\ half-lives and \mee\ have been improved by over an order of magnitude and a factor of 4, respectively, by the current generation of experiments. In the 2021 Neutrinoless Double Beta Decay Portfolio Review~\cite{PortfolioReviewSummary}, DOE-NP identified three experiments ready for execution of the US-led ton-scale \BBz\ decay program, consistent with the 2015 LRP (CUPID, LEGEND-1000, nEXO).  At the same time, the clear consensus is that the quest for \BBz\ decay will not end with the ton-scale program. Discovery of \BBz\ decay at this level will be a tremendous achievement, and should be followed by a campaign of precision measurements in order to understand the decay mechanism. If  Nature is less kind, experiments with sensitivity to the normal ordering of the neutrino masses will be needed. This realization drives a rich R\&D program to prepare for the next-next-generation of discovery experiments. 

\section{Progress Since Last Long Range Plan}
\label{sec:Progress}
The efforts in \BBz\ have made significant progress since the 2015 Long Range Plan~\cite{Geesaman-LRP-2015}. Half life limits now exceed $10^{26}$~yr, ten times longer than those existing in 2015. The constraints on \mee\ now reach near the top of the inverted ordering mass region and, for some isotopes and nuclear matrix element calculations, even extend a bit into this region.
The CUORE~\cite{ADAMS2022CUORE}, EXO-200~\cite{Anton_2019EXO200}, \Gerda~\cite{Agostini_2020GERDA}, KamLAND-Zen~\cite{KamLAND-Zen2022}, \MJ\ \DEM~\cite{Arnquist2022MJDzero}, and  NEXT~\cite{NEXT2nu2022} have established experimental programs demonstrating that experiments at the ton-scale are feasible. \Ltwo~\cite{LEGEND-pCDR} is commissioning and plans physics data taking in late 2022 at LNGS. CUPID-Mo~\cite{CUPID-MOzero, CUPID-MOzero2022} and CUPID-0~\cite{CUPID-0Se} presented final experimental results. SNO+ has measured all of its detector-related 
backgrounds~\cite{SNO:2022trz,Inacio:2022vjt} and shown that it can load up to 3\% Te by mass in its scintillator~\cite{Klein:2022tqr} with acceptable light yield. SuperNEMO~\cite{Arnold_2010} has operated its demonstrator. The progress on isolating and detecting a lone Ba ion within a dense Xe environment was substantially advanced by both the nEXO~\cite{nEXOBaTag2019} and NEXT~\cite{NEXTBaTag2019,NEXTBaTag2021} collaborations.

The breadth of worldwide experimental efforts is truly impressive and demonstrates the community engagement in \BBz\ and excitement about its science. The most notable developments, including operating experiments, R\&D prototypes, and active proposals are listed in Table~\ref{tab:FutureExperiments} in Appendix~\ref{App:AppendixOne}. Following the release of the 2015 LRP, a 2015 subcommittee report to NSAC~\cite{NSAC-BB-Report-2015} listed a number of recommendations related to R\&D challenges for a some of the key US experimental efforts and indicated goals they should accomplish. It is important to emphasize that these goals have now been achieved. The specifics are listed in Appendix~\ref{App:AppendixOne}. 

CUPID~\cite{CUPID-pCDR2019}, \Lthou~\cite{LEGEND-pCDR}, and nEXO~\cite{Adhikari_2021nEXO,nEXO-pCDR}  were selected and reviewed in the 2021 DOE-NP \BBz\ Decay Portfolio Review~\cite{PortfolioReview2022} and all were highly rated and judged to be worth pursuing.
The three experiments are now preparing for the Critical Decision process. This review was followed up by a North America-Europe summit to discuss the future of support for such projects by the various funding sponsors~\cite{SummitReport2021}. Pursuit of these three efforts was endorsed by the APPEC (Astroparticle Physics European Consortium) double beta decay committee~\cite{APPEC-BB2020}, which also noted that NEXT~\cite{Adams2020NEXT} was an exciting future program with a need to demonstrate scalability.

The theoretical description of \BBz\ decay has also made progress. At the time of the previous LRP, calculations of the nuclear matrix elements (\Mz), the main nuclear structure input for the decay, had been performed in a variety of many-body methods that employ empirical interactions, but results varied by factors of 2-3 for the candidate nuclei \cite{Engel_2017}. Since the employed interactions were tailored to specific methods, it was not possible to disentangle uncertainties due to approximations in the many-body method from those due to the parameters of the input interactions. To overcome this challenge, researchers from the Lattice QCD, effective field theory (EFT), and nuclear structure communities initiated a joint effort under the umbrella of the recently concluded DOE Topical Collaboration for Fundamental Symmetries and Neutrinoless Double Beta Decay. This effort led to advances in all involved domains, and it has produced a first wave of \Mz\ values from multiple methods targeting the same candidate nucleus with the same interactions and transition operators. Work is now underway towards a more consistent EFT treatment of interactions and operators, as well as a full statistical uncertainty quantification at all stages of the theoretical description, which will culminate in a next generation of precise \Mz\ values. More details can be found in Sec.~\ref{sec:TheoreticalProgramStatus}, as well as the recent reports \cite{Cirigliano:2022oqy,Cirigliano:2022}.

\section{Current Experimental 
Program}\label{sec:ExperimentalProgramStatus}

The importance of \BBz\ decay is paramount but the experiments are challenging. If an experiment claims evidence for \BBz\ decay, a Nobel Prize winning result, a prompt confirmation would be necessary. The long time frame for construction and operation necessitates that multiple experiments be pursued simultaneously. In addition,  different isotopes, studied with different techniques, will have different experimental uncertainties and different isotopes, with different nuclear matrix elements, will have different theoretical uncertainties. They background conditions of the various experimental approaches are also different. Furthermore results from different isotopes can help unravel the underlying physics that mediates the process.

If lepton number violation is observed in \BBz\ in more than one isotopes the different approaches will provide a robust evidence for discovery.   The US program is now ready to construct experiments at the ton-scale. The following three experiments have undergone a DOE portfolio review and are preparing for 
the Critical Decision process.
\subsection{Next-Generation Ton-scale Experiments}

{\bf CUPID --}The CUORE Upgrade with Particle Identification (CUPID)~\cite{CUPID-pCDR2019} is an upgrade to the operating CUORE experiment aimed at searching for $0\nu\beta\beta$ in $^{100}$Mo in the region of the inverted mass ordering. The proposed CUPID experiment builds on CUORE, the Cryogenic Observatory of Rare Events, at Gran Sasso National Laboratory (LNGS) and leverages the extensive existing cryogenic and technical infrastructure built for CUORE. The baseline design for CUPID features an array of 1596 scintillating crystal bolometers and 1710 light detectors, each instrumented with germanium neutron transmutation doped (NTD) sensors, and organized into 57 towers. While the current design is based on a full complement of enriched Li$_2$MoO$_4$ (LMO) crystals, one key scientific feature of the detector design is its ability to flexibly incorporate multiple isotopes. Bolometric detectors are scalable, allowing gradual, phased deployment. In principle, different double beta decay isotopes may be used in the same infrastructure  in case of a discovery. The total isotopic mass of CUPID will be 240~kg of $^{100}$Mo. The new detector will be installed in an upgraded cryostat at LNGS, taking advantage of the existing infrastructure and facilities developed for use in CUORE. CUPID builds on the success of years of stable operation of the CUORE detector at base temperatures of 10~mK as well as the CUPID-0, CUPID-Mo, and CROSS pathfinder experiments. With light and thermal readout, the estimated background index for CUPID is $<10^{-4}$~c/kg/keV/year. The experiment will have discovery potential in the entire inverted hierarchy region of neutrino masses. We estimate the half-life limit sensitivity (90\%) C.L. at $1.4\times10^{27}$ yr  and the half-life discovery sensitivity ($3\sigma$) of $1.0\times10^{27}$ yr. An expansion to a metric ton of \nuc{100}{Mo} in a larger cryostat or a configuration with multiple cryostats would enable an experiment with sensitivity to the normal hierarchy region. The concept for CUPID-1T, a future ton-scale version of the CUPID concept, is under development. CUPID and CUPID-1T both boast the potential to probe the entire region of the inverted hierarchy of the neutrino mass splitting, and sensitivity to a variety of beyond standard model processes, including symmetry violation and dark matter searches. A future CUPID-1T experiment has the potential to reach into the normal-ordering region of the neutrino mass splitting.

{\bf LEGEND --} The Large Enriched Germanium Experiment for Neutrinoless \BB\ Decay (\Lthou) experiment~\cite{LEGEND-pCDR} utilizes the demonstrated low background and excellent energy performance of high-purity p-type, inverted coax, p-type point contact (ICPC) Ge semiconductor detectors, enriched to more than 90\% in \nuc{76}{Ge}. The background rejection power of ICPC detectors begins with their superb energy resolution, demonstrated to have a full-width at half-maximum (FWHM) resolution of 0.12\% (0.05\% $\sigma$) at \qval. Pulse shape analysis of the signal distinguishes bulk \BBz\ decay energy depositions from surface events and backgrounds from $\gamma$ rays with multiple interaction sites. The granular nature of the Ge detector array allows rejection of background events that span multiple detectors. Finally, background interactions external to the Ge detectors are identified by LAr scintillation light. About 400 ICPC detectors with an average mass of 2.6 kg each are distributed among four 250-kg modules to allow independent operation and phased commissioning. In each module, the detectors are arranged into 14 vertical strings, supported by ultra-clean materials, and read out using ultra-low-background ASIC-based electronics. The detector strings are immersed in radiopure liquid Ar sourced underground and reduced in the \nuc{42}{Ar} isotope. The underground-sourced LAr is contained within an electroformed copper reentrant tube. Each of the four modules is surrounded by LAr sourced from atmospheric Ar, contained within a vacuum-insulated cryostat. The LAr volumes are instrumented with an active veto system comprised of optical fibers read out by Si photomultipliers. The cryostat is enveloped by a water tank providing additional shielding. The baseline design assumes installation in SNOLAB.  The \LEG\ collaboration aims to increase the sensitivity for the $^{76}$Ge \BBz\ decay half-life in a first phase (\Ltwo) to $10^{27}$~yr, and in a second phase (\Lthou) to $10^{28}$~yr, both for setting a 90\% C.L.~half-life limit and for finding evidence for $0\nu\beta\beta$ decay, defined as a 50\% chance for a signal at 3$\sigma$  significance. In \Ltwo\ about 200~kg of Ge detectors will be operated in the existing infrastructure of the GERDA experiment at the LNGS laboratory in Italy. \Ltwo\ is presently commissioning at LNGS and physics data is anticipated beginning in Fall 2022.

{\bf nEXO --} nEXO~\cite{nEXO-pCDR} is based on a Time Projection Chamber (TPC) and the use of five tonnes of liquid xenon (LXe) enriched to 90\% in $^{136}$Xe. The baseline location of the experiment is SNOLAB. This choice is directly derived from the success of EXO-200 and is motivated by the ability of large homogeneous detectors to identify and measure background and signal simultaneously. This approach is taking maximum advantage of the large linear dimensions compared to the mean free path of $\gamma$-radiation. The nEXO TPC consists of a single cylindrical volume of LXe that is instrumented to read out both ionization and scintillation signals in the LXe to obtain $<1\%$ energy resolution~\cite{nEXO_energy_resolution} and strong background rejection. The ionization signal is readout using charge-collection tiles at the top of the TPC while scintillation light is collected with Silicon Photomultipliers (SiPMs) installed around the barrel of the cylinder. The TPC vessel is made from ultra-radiopure custom electroformed copper and is surrounded by a bath of HFE-7000~\cite{HFE}, which acts as a radiopure heat exchange fluid and an efficient $\gamma$-ray shield. The HFE-7000 cryostat is located in an instrumented water tank that serves as a muon veto and additional shielding layer. Information on particle interactions provided by the TPC includes several additional handles to reject backgrounds and improve confidence in a potential discovery. Energy reconstruction, event topology (single vs multi-site interactions), position reconstruction, and scintillation/ionization ratio, are combined using traditional and deep learning tools to effectively discriminate between signal and backgrounds. The nEXO background projections are grounded in existing radioassay data for most component materials and detailed particle tracking and event reconstruction simulations~\cite{Adhikari_2021nEXO}. This approach was validated by EXO-200, where the measured backgrounds closely matched the predictions~\cite{EXO-200_bkgd}. Based on these detailed evaluations, nEXO is projected to reach a 90\% CL sensitivity of $1.35\times10^{28}$ yrs, covering the entire inverted ordering parameter space, along with a significant portion of the normal ordering parameter space, for nearly all values of the nuclear matrix elements. The use of a liquid target has several unique advantages in the case of a discovery. nEXO could directly verify the discovery with a ``blank'' measurement by swapping the enriched xenon with natural/depleted xenon. The enriched target could be reused with a different detector technology, $e.g.,$ a discovery with nEXO may be followed by an investigation of energy and angular correlations in a gas TPC.

\subsection{Ongoing and Future Experiments}

In addition to the ton-scale experiments a number of complementary efforts with significant US contributions are advancing the field. Some of these experiments already have data and some will be taking data soon, with results expected before the start of the ton-scale program.  Some also use isotopes different from those that will be used in the ton-scale program, and many of these experiments provide paths toward much larger detectors that could go beyond the ton-scale and start to probe the normal hierarchy region.  
Should $0\nu\beta\beta$ decay be discovered at the ton-scale, techniques being developed in these experiments may provide ways of unraveling the $0\nu\beta\beta$ 
mechanism. 
We describe each of these below.

{\bf NEXT --}The Neutrino Experiment with a Xenon TPC (NEXT)~\cite{Adams2020NEXT} is a sequence of high pressure xenon gas time projection chambers (HPGXeTPC) targeting ultra-low background indices in high pressure xenon gas.  The virtues of HPGXeTPC technology include its excellent energy resolution relative to other xenon-based techniques (intrinsic resolution FWHM/E=$\sim$0.3\% at the Q-value of \BBz); topological imaging capabilities that can distinguish two-electron signals from single-electron backgrounds; a combination of scalability and modularity; and potential to implement barium daughter tagging in situ via single molecule fluorescence imaging (SMFI). The 10~kg NEXT-White  demonstrator at the LSC Laboratory has recently completed its run plan, demonstrating energy resolution below 1\% FWHM in xenon ($\sigma_E/E\sim0.4$\%); proving the background rejection power of topological discrimination using both traditional methods and deep neural networks; validating the NEXT background model; and measuring $2\nu\beta\beta$ based on event-by-event topological identification and direct subtraction between enriched and depleted xenon.  The NEXT-100 phase is now under construction with operation scheduled to begin in 2023. An expression of interest for a first ton-scale module (NEXT-HD) with incremental improvements over NEXT-100 has been submitted to the LSC laboratory, with projected start of construction in 2026.  The NEXT Collaboration is also pursuing a more disruptive approach based on the efficient detection of the Ba$^{2+}$ ion produced in the double beta decay of \nuc{136}{Xe} using single-molecule fluorescence imaging (SMFI). NEXT R\&D on barium tagging has seen major technological advances in areas including in single molecule microscopy, organic chemosensor development, characterization of ion capture in dry conditions and vacuum, and transport techniques.  In particular, a US-based barium tagging demonstrator phase called NEXT-CRAB is under development that will combine optical TPC readout and barium detection subsystems with the goal of characterizing the efficiency and background levels using both barium ion beams and subsequently the daughters from topologically identified $^{136}$Xe \BBt\ in xenon gas.  During the coming LRP period, the NEXT Collaboration aims to demonstrate the viability of an ultra-low background, beyond-the-ton-scale detector concept to reach toward the normal mass ordering region of \BBz\ parameter space.  The combination of tracking individual electrons and barium tagging may also provide unique information to confirm a discovery and disentangle the dominant decay mechanism, should a \BBz\ signal be observed.

{\bf SNO+ --} SNO+ uses a ``loaded'' liquid scintillator technique~\cite{Inacio:2022vjt}. {\it Natural} tellurium, with a 34\% isotopic abundance of the $\beta\beta$ isotope $^{130}$Te, will be loaded into 780 tonnes of LAB-PPO scintillator, viewed by the original Sudbury Neutrino Observatory (SNO) detector. The goal is a loading of up to 3\% Te by mass in the scintillator, with light yield good enough that backgrounds from the $2\nu\beta\beta$ decay are manageable. External backgrounds, mainly $\gamma$ rays  from the acrylic vessel holding the scintillator, the light-water shield, and the photomultiplier tube array, are significantly suppressed by the detector's very large size and the imposition of a strict fiducial volume cut.  Backgrounds from internal radioactivity, primarily U and Th-chain decays of $^{214}$BiPo and $^{212}$BiPo, are mitigated by the coincidence of the bismuth $\beta$ and polonium $\alpha$, as well as the separation based on the different time profiles of the $\beta$s and $\alpha$s.  Multi-site $\gamma$ events, like those from cosmogenic isotopes like $^{60}$Co, can be separated from a putative $0\nu$ signal based on the spread in their PMT time residuals, and underground purification of the Te is expected to render these backgrounds negligible in any case. An initial test deployment of up to 0.5\% Te by mass is planned in the next few years, with expected sensitivities of $m_{\beta\beta} < 30-180$~meV after three years of data taking.  The 0.5\% Te has been underground and cooling down for several years, adding to the suppression of any cosmogenic isotopes, even in advance of purification.  SNO+ also has a broad program of other neutrino physics, including reactor antineutrino measurements of $\Delta m^2_{12}$, low-energy $^8$B solar neutrinos, geo-neutrinos, and supernova and other astrophysical neutrino sources. At a 3\% loading, given current {\it in-situ} background measurements, SNO+ sensitivity is expected to be $m_{\beta\beta}<15-40$~meV after ten years of running.  A successful deployment of SNO+ will not only provide the most stringent limits on Te $0\nu\beta\beta$ decay (already at 0.5\%), and reach the bottom of the inverted hierarchy region (with 3\%), it will serve as a demonstrator of a much larger and affordable experiment that could probe the normal hierarchy region, like the proposed \textsc{Theia} experiment.

{\bf KamLAND-Zen --}The KamLAND-Zen experiment searches for \BBz\ decay in liquid scintillator loaded to 3\% by weight with Xe gas enriched to 90\% in $^{136}$Xe. The Xe-loaded LS is deployed in a thin nylon balloon at the center of the ultra-low background KamLAND detector. In two major phases, deploying up to 380~kg (KamLAND-Zen 400) and 750~kg (KamLAND-Zen 800) of enriched Xe, KamLAND-Zen reported world-leading half-life limits. The most recent limit of $T_{1/2} > 2.3 \times 10^{26}$~yr probes the top of the Inverted Ordering region for the first time for at least one nuclear matrix element calculation method~\cite{KamLAND-Zen2022}. The background at $Q_{\beta\beta}$ is dominated by long-lived xenon spallation products and by the 2$\nu\beta\beta$ tail. An upgrade of the detector, dubbed KamLAND2-Zen~\cite{Shirai2017}, aims for a factor-of-two improvement in energy resolution due to a combination of improved liquid scintillator and a refurbishment of the PMT array, using higher-quantum-efficiency PMTs instrumented with light concentrators. An upgrade to the detector electronics aims to further reduce spallation backgrounds. With 1~ton of enriched Xe, KamLAND2-Zen is expected to reach half-life sensitivities in excess of 10$^{27}$~yr.

{\bf \textsc{Theia} --} The \textsc{Theia} experiment concept leverages both Cherenkov and scintillation light in a so-called ``hybrid'' neutrino detector in order to achieve unprecedented levels of event and particle identification capabilities in a low-threshold, low-background detector~\cite{Askins:2019oqj}.  Directional information from the Cherenkov light will allow rejection of the solar neutrinos that become a dominant background as experiments are scaled to larger sizes.  A combination of the scintillation time profile, Cherenkov / scintillation ratio, and multi-site discrimination allow additional background rejection, resulting in sensitivities at the meV scale, for conservative levels of loading of either natural Te or Xe.  A suite of prototype detectors are being constructed to demonstrate the capabilities of this technology, leveraging a decade of bench-top scale development of novel scintillators, fast photon detectors, and spectral sorting~\cite{Caravaca:2016fjg,CHESS2020,Biller:2020uoi,Guo:2017nnr,Lyashenko:2019tdj,osti_1564252,Klein:2022tqr,Kaptanoglu:2017jxo,Kaptanoglu:2019gtg,CHESS2016,CHESS2020,Kaptanoglu:2019gtg,Gruszko:2018gzr,Kaptanoglu:2021prv,mainz}.  
A four hundred kg deployment of WbLS in the Booster Neutrino Beam at ANNIE~\cite{sandi} will allow the first demonstration of high-energy neutrino event reconstruction with this technology. Brookhaven National Laboratory (BNL) has a ton-scale deployment ongoing, allowing performance and stability testing with cosmogenic muons~\cite{wbls_recent}. A 30-ton vessel is also planned at BNL, which will be the first large-scale deployment. The \textsc{Eos} detector is being constructed at Berkeley, with a 4-ton fiducial volume, to demonstrate the event reconstruction capabilities, and the impact of different detector configuration choices~\cite{eos}.
A proposal is underway to construct \textsc{Theia} at the Long Baseline Neutrino Facility in South Dakota, a deep underground location that would offer a broad program of low energy physics -- including percent-level precision on a measurement of CNO neutrinos, probes of the MSW transition region, supernova and solar neutrinos, geoneutrinos, and sensitivity into the normal ordering region for a \BBz\ search -- as well as CP violation sensitivity from exposure to the high-energy neutrino beam from FermiLab.
The impact of this technology to this rich program of physics is explored in papers such as~\cite{Aberle:2013jba,Bonventre:2018hyd,Askins:2019oqj,TheiaAntinu,Land:2020oiz,TheiaDSNB,Elagin:2016zgp}.

{\bf SuperNEMO --} The SuperNEMO Experiment continues to demonstrate a technique originated by the NEMO-3 Experiment~\cite{NEMO-3-NIM} of using thin isotopic foils surrounded by a wire drift chamber for 3D particle tracking in a 25\,G magnetic field, and scintillator blocks to measure the electron energy. The detector reconstructed the topology, energy, and timing features of nuclear decays. This unique technique provides several observables for each registered event offering a powerful means to identify double beta decays and to reject background processes. This method allows some freedom of choice of the isotopic sources but suffers from necessity of large detector footprint although there are ideas to overcome this shortcoming. The SuperNEMO Demonstrator Module is currently being commissioned in the Laboratoire Souterrain de Modane in the 4,800\,m.w.e. deep  Fr\'ejus tunnel in France. 
The physics program of the SuperNEMO Demonstrator Module consists of precision measurements of the \BBt\ decay mode to constrain nuclear and BSM physics, as well as the best limits on \BBz, for the isotope \nuc{82}{Se}. Studies are ongoing how to extend the unique tracker-calorimeter technique to explore the mechanism of \BBz\ should it be discovered in another experiment for another isotope.

The richness and excitement of the worldwide \BBz\ decay program is summarized in Table~\ref{tab:FutureExperiments}, and the best limits to date are given in Table~\ref{tab:PastExperiments}.

\begin{table*}[ht]
\caption{A list of the best  \BBz\ decay \Tz\ limits at 90\% confidence level for several isotopes. The \mee\ limits are those quoted by the authors using the matrix elements of their choice.}
\begin{center}
\begin{tabular}{|c|c|c|c|c|}
\hline
Isotope                &  Technique                   			  & \Tz\                               	  & \mee\ (eV)       		& Year Published  \\
\hline
\hline
\nuc{48}{Ca}     & CaF$_2$ scint. crystals           		  &$>5.8 \times 10^{22}$ y          &   $<$3.5-22   		& 2008~\cite{Umehara2008}\\
\nuc{76}{Ge}   & \nuc{76}{Ge} detectors                   		   	&    $>1.8 \times 10^{26}$ y        &      $<$0.079-0.180     	& 2020~\cite{Agostini_2020GERDA} \\
\nuc{82}{Se}    &Zn\nuc{82}{Se} bolometers               	& $>4.6 \times 10^{24}$ y              & $<$0.263-0.545       	 &  2022~\cite{CUPID-0Se}\\
\nuc{96}{Zr}    & Thin metal foil within TPC                & $>9.2 \times 10^{21}$ y                  & $<$3.9 - 19.5             &2009~\cite{Argyriades_2010}   \\
\nuc{100}{Mo}   &Li$_2$\nuc{100}{Mo}O$_4$ bolometers              	 & $>1.8 \times 10^{24}$ y             & $< $0.28-0.49        	 &  2022~\cite{CUPID-MOzero2022}\\
\nuc{116}{Cd}   &\nuc{116}{Cd}WO$_4$ scint. crystals 	& $>2.2 \times 10^{23}$ y              & $<$1.0-1.7       		& 2018~\cite{Barabash2018}\\
\nuc{128}{Te}    & TeO$_2$ bolometers                           		& $>3.6 \times 10^{24}$ y               & $<$1.5-4.0         	& 2022~\cite{CUORE:2022piu} \\
\nuc{130}{Te}    & TeO$_2$ bolometers                     		& $>2.2\times 10^{25}$ y              & $<$0.090-0.305       	 &  2022~\cite{CUORE-0Te}  \\
\nuc{136}{Xe}   &  Liquid Xe scintillators  					& $>2.3 \times 10^{26}$ y 		&  $<$0.036-0.156 		& 2022~\cite{KamLAND-Zen2022}    \\
\nuc{150}{Nd}   &   Thin metal foil within TPC    			&$>2 \times 10^{22}$ y               & 1.6-5.3                 		 & 2016~\cite{Arnold2016} \\
\hline
\end{tabular}
\end{center}
\label{tab:PastExperiments}
\end{table*}%

\begin{figure*}[ht]
 \centering
\includegraphics[width=18cm]{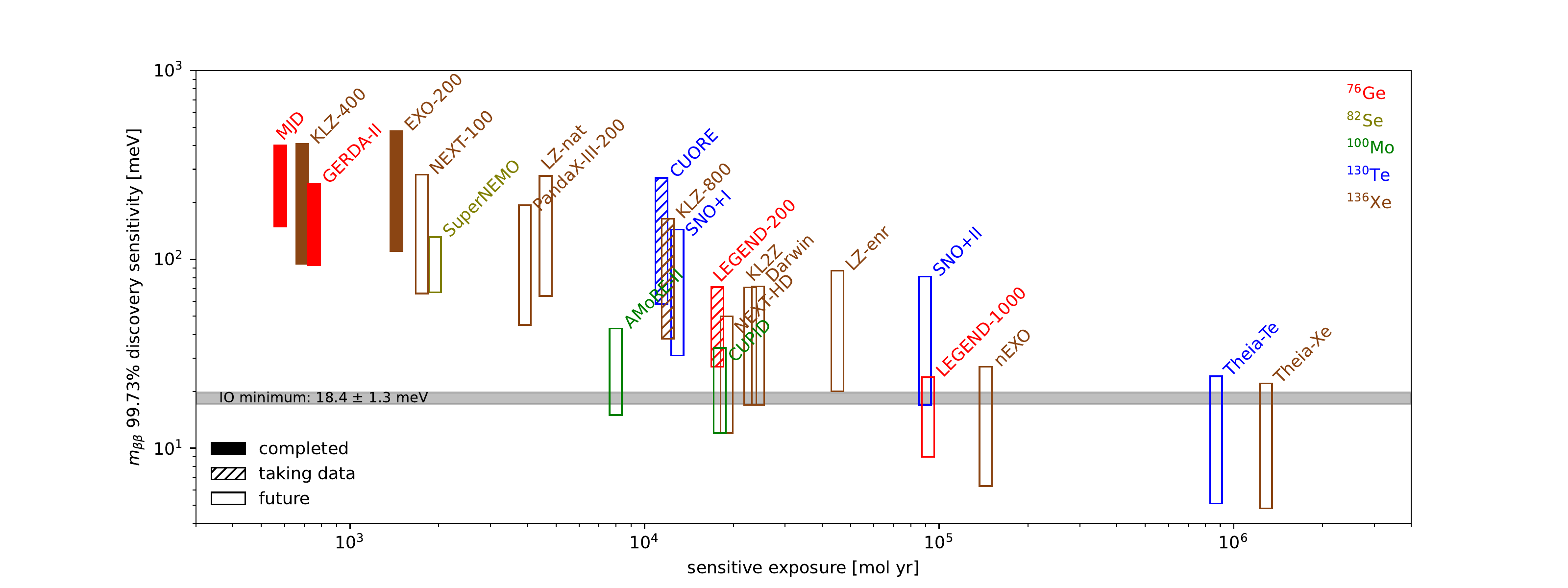} \par
\includegraphics[width=8cm]{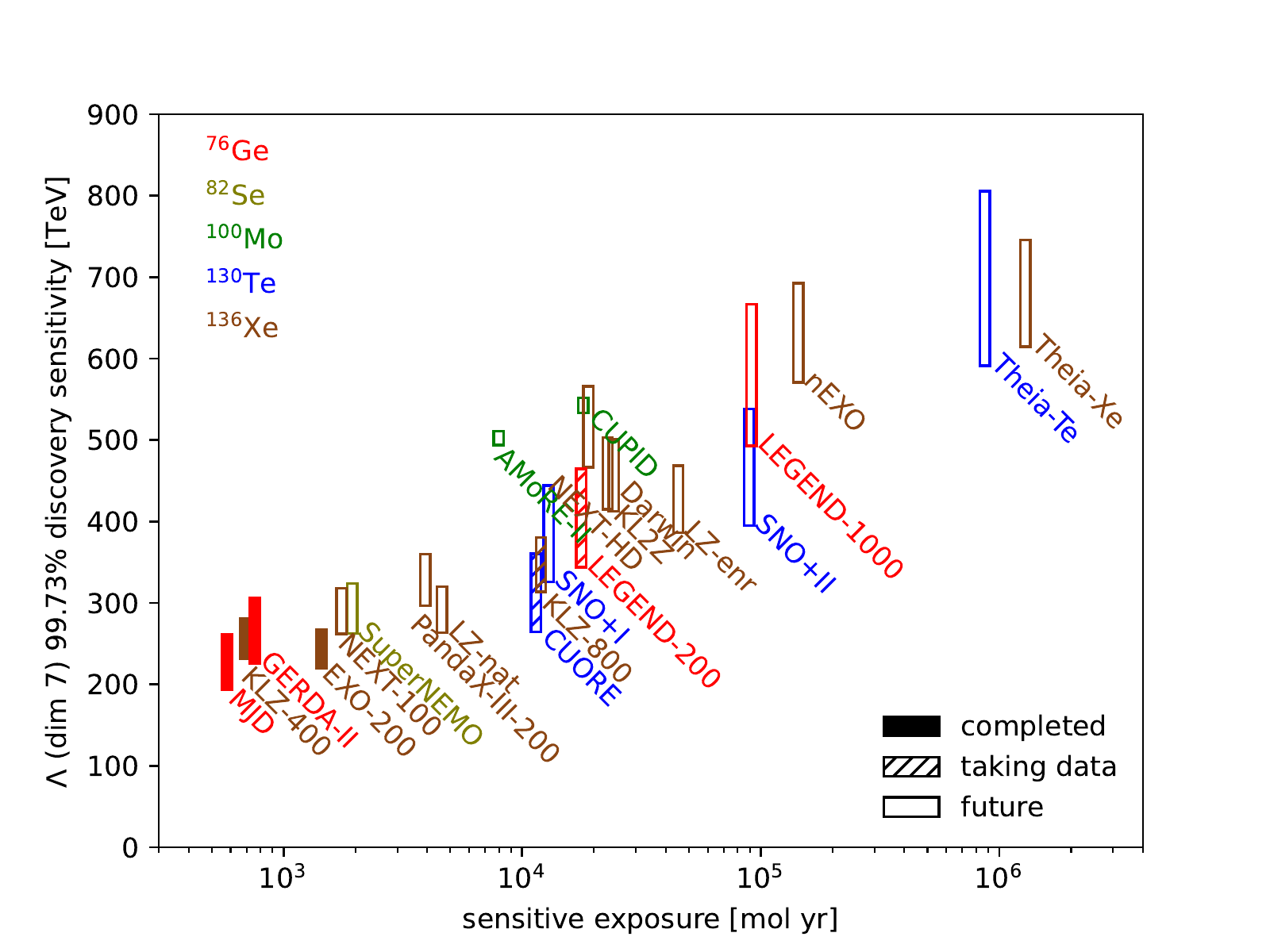}
\includegraphics[width=8cm]{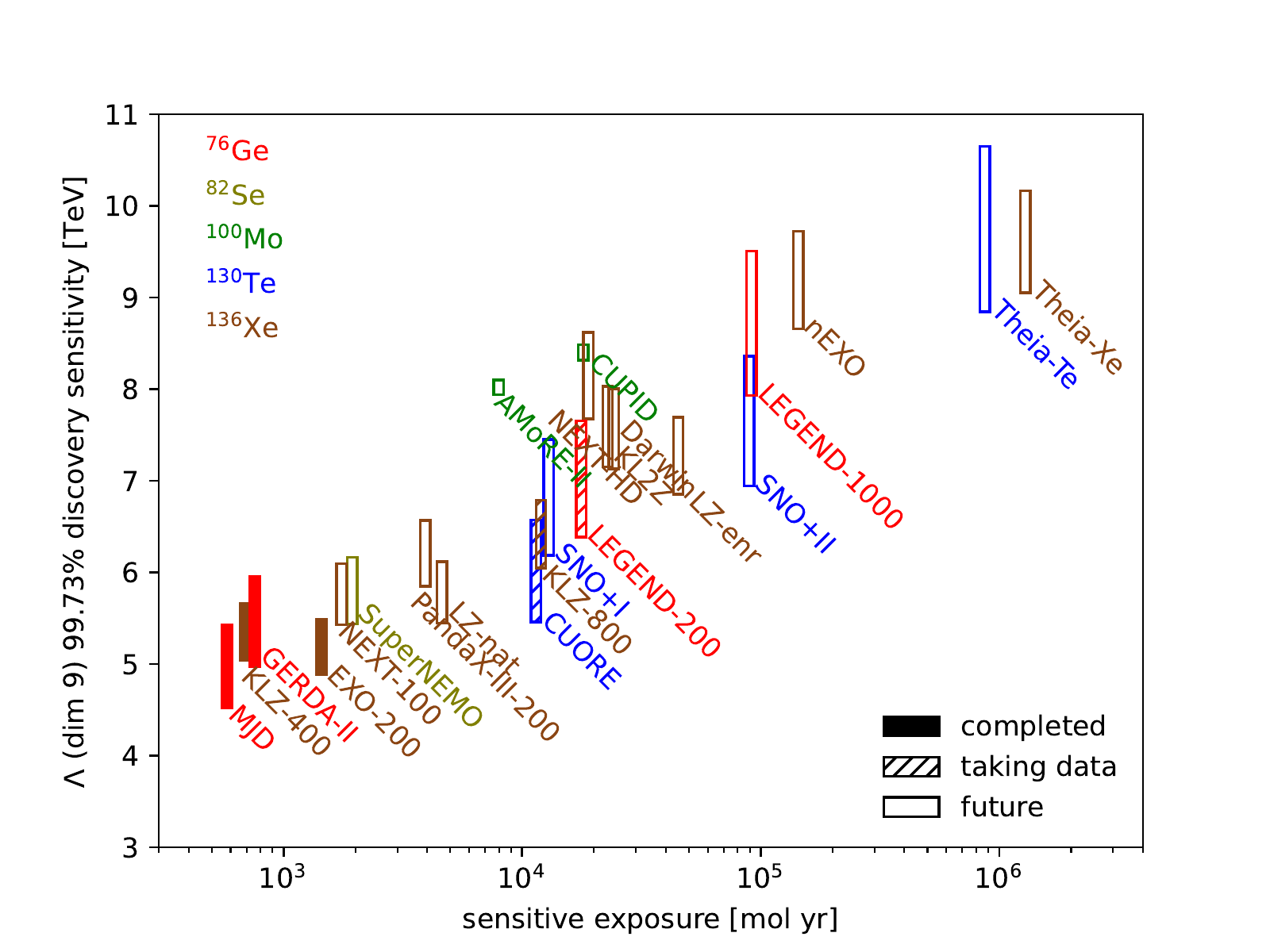}
 \caption{
Discovery sensitivities of current- and next-generation \BBz-decay experiments for 
various mechanisms of lepton number violation, 
dominated by effective operators of dimension 5 (top panel), i.e., light neutrino exchange, and of
dimension 7 (lower left panel) and dimension 9 (lower right panel). 
Values of \mee\ larger than the marked values are tested at higher CL. 
Values of $\Lambda$ smaller than the marked values are tested at higher CL. 
At dimension 7 and 9, 
we  show  the reach for a single operator, the one that 
is least suppressed by chiral and electroweak scales~\cite{Cirigliano:2018yza}.
The size of the bar indicates the spread of the nuclear matrix elements (NMEs) and should be understood as a conservative range, not a standard deviation.
The nuclear matrix elements are taken from~\cite{Menendez:2017fdf,Horoi:2015tkc,Coraggio:2020hwx,Coraggio:2022vgy,Mustonen:2013zu,Hyvarinen:2015bda,Simkovic:2018hiq,Fang:2018tui,Terasaki:2020ndc,Rodriguez:2010mn,LopezVaquero:2013yji,Song:2017ktj,Barea:2015kwa,Deppisch:2020ztt,Jiao:2017opc}. Some matrix elements~\cite{Coraggio:2020hwx,Coraggio:2022vgy} include an initial estimate of quenching mechanisms that require further work.
The IO minimum is taken from~\cite{Agostini:2021kba}. Figure adapted from Ref.~\cite{Agostini2022MatterDiscover}. 
}
\label{fig:ExperimentComparision}
\end{figure*}

Figure.~\ref{fig:ExperimentComparision} shows a comparison of the sensitivities of various recent and future \BBz\ efforts~\cite{Agostini2022MatterDiscover}.

\subsection{Beyond Next-Generation Ton-Scale Experiments} 

The field of neutrinoless double beta decay will continue beyond the current, ton-scale experiments. Should \BBz\ decay be discovered and confirmed by several ton-scale experiments, a Nobel-Prize-worthy result, the next step would be to identify the mechanism behind LNV. Sensitivity to different models of new LNV physics could be achieved by precision measurements of the half-lives of different isotopes, and by measuring the event topology, such as energy and angular distributions of the decay electrons. 

If \BBz\ decay is not observed with the half-lives consistent with the inverted ordering of neutrino masses, increasing the scale and sensitivity of the experiments will be of paramount importance.
Such experiments will require even larger isotopic masses, typically at or above the ton scale and very low, ideally negligible, backgrounds. Reconstructing topology of the events would be important. 
Several concepts for experiments with the sensitivity below the inverted mass ordering scale and well into the normal hierarchy region exist, as discussed above. It is important to support R\&D to identify the most promising technologies over the next decade or so in order to be ready to mount the ambitious next-next-generation experiments by the time the ton-scale experiments complete their operations. 

\section{Status of Theoretical Program}\label{sec:TheoreticalProgramStatus}

The interpretation of \BBz\ decay experiments and, in case of discovery, the identification of the underlying mechanism behind a signal require an ambitious theoretical program, with several interconnected components, ranging from lepton number violating (LNV) phenomenology to the calculation of the relevant hadronic and nuclear matrix elements with quantified uncertainties. 
The breadth and depth of this program is captured by the recent reports found in Refs.~\cite{Cirigliano:2022oqy}  and~\cite{Cirigliano:2022}, which also offer a detailed bibliography. Here we summarize some of the salient features. 

{\bf LNV phenomenology:}  In this area the need exists to further explore models of LNV and neutrino mass that go beyond the high-scale see-saw  paradigm, and test them against the results of current and future \BBz\ decay experiments and other probes across energy scales. These probes include other low-energy neutrino experiments,  high energy colliders,  astrophysics,  and cosmology (e.g. connection of TeV-scale LNV with leptogenesis mechanisms).  
Recent highlights~\cite{Li:2020flq,Li:2021fvw,Harz:2021psp,Graesser:2022nkv} 
and future prospects are discussed in detail in Ref.~\cite{Cirigliano:2022oqy}. 

{\bf Hadronic and nuclear matrix elements:} In this broad program,  the goal is to compute \BBz\ decay rates with minimal model dependence and quantified theoretical uncertainty by advancing progress in particle and nuclear effective field theories (EFTs), lattice quantum chromodynamics (QCD), and \emph{ab-initio} nuclear-many-body methods.

At the time of the 2015 LRP, nuclear matrix elements from a wide variety of many-body approaches --- the QRPA, the Shell Model, DFT and the IBM --- had been computed, but results for important nuclei varied by factors 2-3, with no guarantee that the correct matrix elements were within the spread (see \cite{Engel_2017} and a more recent update in \cite{Agostini2022MatterDiscover}).  It is difficult to assess the quality of any of these calculations and to compare them because, for example, they each use empirical interactions that are not appropriate for other methods, and they each make ad-hoc assumptions about the effects of short-range correlation effects on the transition. To tackle such problems, the Lattice-QCD, EFT and nuclear-structure communities launched a collaborative effort to develop a consistent, systematically improvable framework for ab initio matrix elements: EFT to specify the form of the decay operator, a combination of lattice QCD, modeling, and fitting to determine the constants that multiply particular terms in the operator, and \emph{ab initio} nuclear-structure theory to solve the nuclear many-body problem and compute the final matrix element.  Encouraging progress has been made in the last few years on the EFT and lattice QCD (LQCD) aspects of the problem. 
The EFT framework for \BBz\ has been   developed
for the light Majorana neutrino exchange~\cite{Cirigliano:2017tvr,Cirigliano:2018hja,Cirigliano:2019vdj}
and the TeV-scale mechanisms ~\cite{Prezeau:2003xn,Cirigliano:2017djv,Cirigliano:2018yza}, 
with the inclusion of sterile neutrinos~\cite{Dekens:2020ttz}.
Progress has been made in LQCD 
for the  $\pi^- \pi^- \to e e$ process~\cite{Nicholson:2018mwc,Feng:2018pdq,Tuo:2019bue,Detmold:2022jwu} 
and towards two-nucleon amplitudes~\cite{Feng:2020nqj,Davoudi:2020gxs}. 
The error in each of these steps (EFT truncation, effective couplings, and nuclear structure) can in principle be quantified, and will eventually lead to a matrix element with a meaningful uncertainty.

\begin{figure*}[ht]
 \centering
 \includegraphics[width=0.9\textwidth]{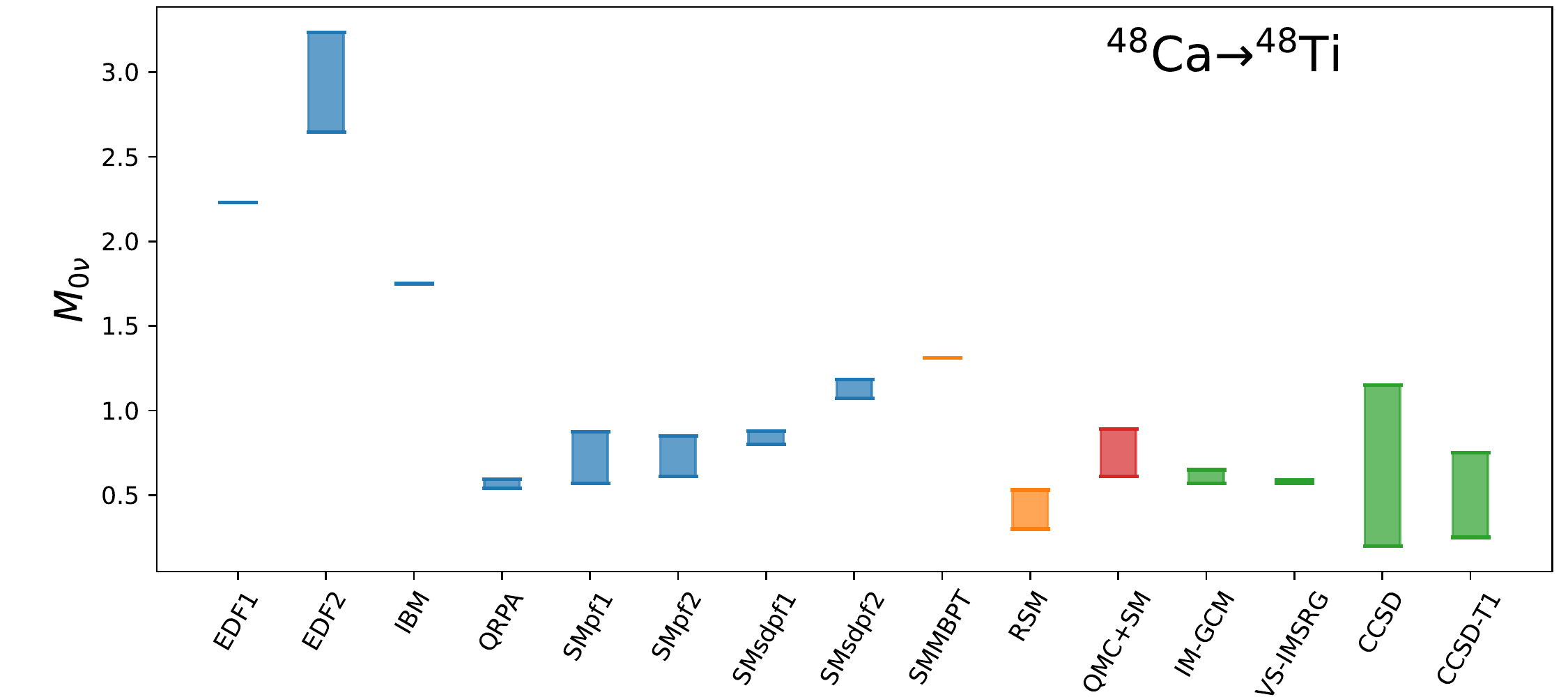}
 \caption{
 Nuclear matrix elements (\Mz) for the \BBz\ 
 decay $\nucl{Ca}{48}\rightarrow\nucl{Ti}{48}$. Blue symbols are results from a variety of many-body methods with empirical or semi-empirical tailored interactions, as well as a mixture of schemes for treating short-range correlations. Orange symbols aim at a more systematic treatment of interactions and correlations via Many-Body Perturbation Theory. The red NME is from a combined shell-model and QMC analysis that primarily focuses on the short-range correlations. Green symbols indicate results from complementary, fully non-perturbative \emph{ab initio} calculations, all using the same interaction and transition operators as input. 
 Shaded areas represent only the sensitivity of the matrix element to certain parameters of the calculation, not complete theoretical error bars.  Figure adapted from the recent report \cite{Cirigliano:2022}, courtesy of R. Stroberg --- for additional details, see original reference and references therein. 
 }
 \label{fig:Ca48NME}

\end{figure*}

The upper panel of Figure \ref{fig:Ca48NME} illustrates a first-generation application of this framework to the matrix element for the decay $\nucl{Ca}{48}\to\nucl{Ti}{48}$. Results in blue come from a variety of methods that use empirical or semi-empirical interactions and different ad-hoc prescriptions for short-range correlation effects.  By contrast, the ab initio results in green employ the same interaction and transition operator as input for several \emph{complementary} many-body methods (see \cite{Cirigliano:2022} and references therein for details). The shaded areas are not full uncertainty estimates; they indicate only the sensitivity of the matrix element to certain parameters in the calculation and neglect, e.g., EFT uncertainties that would make them larger.  The uncertainty estimates for phenomenological results in blue cannot easily be improved, but those for the ab initio calculations can, as we discuss below.  

An important result from the effort to develop consistent EFT interactions and transition operators is the discovery that the exchange of high-momentum virtual neutrinos between nucleons contributes non-negligibly to the decay, and in ways that cannot simply be modeled by nucleon form factors or short-range correlations between nucleons.  In the systematic EFT approach, this physics manifests itself as an additional term in the \BBz\ decay operator with zero range~\cite{Cirigliano:2018hja}.
Recently, a  calculation  of the $nn \to pp$ amplitude near threshold  has been carried out
with dispersion-theory techniques truncated to the elastic two-nucleon channel~\cite{Cirigliano:2021hp,Cirigliano:2021gz}.
This has allowed nuclear-structure practitioners~\cite{Wirth:2021vu} 
to determine the coefficient of the contact term and implement it in calculations, where it leads to a non-negligible and robust enhancement of the NMEs.

Nuclear-structure theory itself has seen a number of important developments.  \emph{Ab inito} techniques seem to have almost fully explained the systematic quenching of single-$\beta$ decay rates that goes by the moniker ``$g_A$ quenching.''  A combination of correlations that have escaped phenomenological models (such as the shell model and the QRPA) and two-body weak currents (corresponding to meson exchange during the decay) are responsible.
Both mechanisms have been investigated within \BBz\ decay.  Correlations reduce those matrix elements as well, and the effects of two-body currents are still not fully quantified.  The pieces we can compute have mostly small effects, but another zero-range term with an unknown coefficient has yet to be assessed.  Theory is now moving beyond the matrix elements summarized in Fig.~\ref{fig:Ca48NME}; matrix elements for $^{76}$Ge and $^{82}$Se are starting to come in as well.  The new matrix elements, especially in Ge, are smaller than those produced by phenomenological models, but just how much smaller they are is an open question because theoretical uncertainty is still significant.

What are the next steps~\cite{Cirigliano:2022} in the theory program?

\begin{itemize}
    \item First, the \emph{ab initio} methods must continue to improve. These approaches are defined through truncation schemes that provide systematic convergence to an exact result. However, improved truncations also imply significantly increased computational costs; this is the main reason current truncations are more restrictive than we would like. 

    \item In addition, the  structure couplings  of the EFT decay operators must be fully specified, including those that appear in sub-leading order in the nuclear EFT, 
such as in the two-body currents.  
This program can be carried out by studying systems of two and three nucleons 
through a combination of EFT,  dispersive methods, and ultimately lattice QCD.


\item   Hadronic and nuclear matrix elements relevant for TeV-scale LNV mechanisms 
require  more dedicated study.  The nuclear-structure community has thus far focused almost exclusively on 
light-neutrino exchange. 

\item 
The community must carry out a robust uncertainty-quantification program, as laid out in detail in Ref. \cite{Cirigliano:2022}.  
This in itself will require several steps: 

\begin{itemize}

   \item  Quantifying the EFT truncation error  by performing nuclear calculations with interactions and transition operators 
   truncated at different orders.

    \item  Developing ``emulators'' for the \emph{ab initio} methods --- surrogates that can approximate the results of the method they emulate in much less time.  That step will allow the community to examine correlations between observables, vary Bayesian priors, construct posteriors, etc.  Emulators exist for some methods 
    but for others their development will require more work.  

    \item Deciding how to combine the predictions of various methods to produce a single matrix element with an uncertainty that reflects the community's confidence in each method.  Here an analysis of the ability of methods to reproduce observables correlated with \Mz\ is essential.  Carrying it out means first quantifying the correlations, then examining the predictions of each model, which can be ``scored'' so that one can decide how much weight to give its predictions for \Mz.
\end{itemize}

\end{itemize}

Carrying out the  multi-pronged theoretical program outlined in this section  is an integral part of a 
successful US-led \BBz\ decay science campaign. 
Seeing it through will require more resources than we have at present, in the form both of person power and of computational power.

\section{Workforce Development, Diversity, Equity, and Inclusion }

Experimental and theoretical collaborations in neutrino physics represent diverse and
multi-cultural teams of scientists from across the world. The variety
of detector technologies, size of collaborations, and experimental backgrounds
provide unique opportunities for creative scientific growth of our students and postdocs and excellent training for the
next generation of nuclear scientists. Seizing these opportunities
while providing a nurturing, supportive, respectful, inclusive, and accessible
environment is an important part of our mission. Collaborations in double beta decay science can play an important role in the endeavour.

Double beta decay experiments and associated R\&D efforts offer a variety of opportunities and pathways for students and pathways from a range of backgrounds. 
In particular, modern science experiments and large and diverse scientific collaborations that operate them, enable opportunities at institutions that do not traditionally have access to big science. Including 
undergraduate-only institutions, emerging research institutions,  and
minority-serving institutions in 
the \BBz\ decay scientific collaborations is an opportunity to grow our field and attract a broader cross section of the population into nuclear science. Recent
initiatives like FAIR~\cite{FAIR} and RENEW~\cite{RENEW} in the Office of Science and similar
initiatives at the NSF and at institutions are excellent opportunities for the \BBz\ decay collaborations. Several institutions already play an important role in these programs and offer in training and mentoring for the next-generation of undergraduates, postbac students and graduate students.

The long timescales and distributed nature of the current and
next-generation of \BBz\ decay experiments do present challenges in
workforce development.
These create obstacles
to a fully inclusive and diverse environment, and we must strive
to mitigate them to the greatest extent possible. In particular,
while participating in construction and operation of the
next-generation projects, the \BBz\ decay collaborations have the responsibility to provide early
career scientists with opportunities for scientific productivity, 
 creative development of new ideas through both targeted and blue-sky
R\&D  efforts, and for visibility and leadership both within their
collaborations and institutions as well as on the global stage. 
Providing these opportunities require adequate material, financial, and administrative support for the research programs. 

Several aspects of our work that we take as given add pressure on the work-life balance, family
responsibilities, and mental health of early-career
scientists. These include 
collaborations spread across many time zones, necessity of international
travel and deployment at remote sites away from the supportive
environment of the home institutions, unregulated work hours, and
project duration exceeding typical timescales for a PhD or postdoc
appointments. These pressures affect the disadvantaged and
vulnerable segments of our scientific population most and contribute to
career impediments. 
The field and the individual collaborations needs to be mindful of these pressures.
Remote and hybrid meetings need
to be carefully designed to allow full and inclusive participation,
including by persons with disabilities.

The \BBz\ decay collaborations have taken an active approach in developing explicit policies that uphold the core
principles and values of supporting diversity, inclusion, and
equity. These are consistent with guidance from the
APS and the funding agencies, but may need to go further. Procedures to handle cases of
harassment, intimidation, and micro-aggressions need to be implemented carefully and with commitment to fairness, due process, and protection of the survivors. This is particularly
sensitive for multi-cultural international collaborations. 

A summary of the HEP Climate, prepared for the Snowmass Community Engagement Frontier report~\cite{Hansen:2022xdy}, summarizes most issues of work force development, diversity, inclusion, and equity for the particle physics community. Most of these issues are common to nuclear physics, and the recommendations are broadly applicable. Initiatives across the divisions of the APS and across the funding
agencies to promote diversity, equity, and inclusion, to ensure fair,
accessible, and equitable work environment, and to empower early career
researchers are being pursued. The \BBz\ decay collaborations can take a leading role in these efforts. Such broad efforts would reinforce the
common set of values and expectations across the APS, and would allow
sharing of ideas, learning from best practices, and help move all
fields forward.

\newpage
\section{Conclusion}\label{sec:Conclusion}
If \BBz\ is established, the implications will be profound. We will immediately know that neutrinos are their own antiparticles, that lepton number is not conserved, and hence will have direct evidence of physics beyond the Standard Model.  These conclusions would be independent from theoretical arguments.   The discovery would also offer one necessary condition and a plausible explanation for the observed matter-antimatter asymmetry in the universe.
Measurement of the rate would probe the neutrino mass scale, provide a terrestrial constraint on the standard
cosmological model, and yield insights into mass generation. Such extraordinary results require correspondingly convincing evidence.
No single experiment based on a particular isotope will be sufficient.
What is required is independent observations in multiple isotopes, with different
experimental methods and systematics.
Since the 2015 Long Range Plan, the US nuclear physics community, in collaboration with our international partners, 
has developed a new generation of ton-scale experiments capable of probing the inverted ordering parameter space and answering this challenge.
Three international experiments, CUPID, LEGEND, and nEXO, based on three different isotopes and technologies, 
all with significant US involvement, were deemed ready to
proceed following a comprehensive DOE Portfolio Review carried out during the summer of 2021.
Accordingly, and consistent with the 2015 Long Range Plan recommendation II,
recommending the timely development and deployment of a US-led, jointly with international partners, ton-scale neutrinoless double beta decay experiment,
we propose the following recommendation for the 2023 Long Range Plan: \\

{\bf We recommend timely construction of ton-scale neutrinoless double beta decay experiments using multiple isotopes.}\\

We note that mounting three experiments with three different isotopes can only be accomplished with both significant US involvement and support
as well as significant collaboration with and contributions from international partners.   

These efforts must be in conjunction with support for a healthy nuclear theory program which is vital for providing the needed physics underpinning, interpretation, and planning base for this effort. Given the importance and impact of discovering \BBz\ it is essential that the
community continues to actively explore and develop improved experimental approaches beyond the ton scale experiments. Such R\&D will be essential for either interpreting the discovery in terms of the underlying physics, or reaching beyond the inverted mass ordering, should that be needed.
Correspondingly, we propose the accompanying recommendation for the 2023 Long Range Plan \\


{\bf We recommend support for a robust research program in neutrinoless double beta decay that includes the ongoing efforts in theory and experiment
as well as a diverse R\&D program exploring multiple promising isotopes and technologies with sensitivity beyond the inverted mass ordering.}\\

\newpage
\appendix
\renewcommand{\thesection}{\Alph{section}}

\section{Progress Since 2015 NSAC Report}\label{App:AppendixOne}

A 2015 subcommittee report to NSAC~\cite{NSAC-BB-Report-2015} listed a number of recommendations related to R\&D challenges for a number of the key US experimental efforts and indicated goals they should accomplish. These goals have been achieved and are listed here. 

\begin{enumerate}
\item SNO+ background model is now understood~\cite{SNO:2022trz,Inacio:2022vjt}
\item SNO+ has found a path to 3\% loading of Te~\cite{Klein:2022tqr}.
\item SNO+ has demonstrated the required 200 photoelectrons/MeV light collection requirement~\cite{Inacio:2022vjt}.
\item NEXT has demonstrated a projected background in a ton-scale project below 1~\cpFty~\cite{adams2021sensitivity,novella2022measurement}.
\item NEXT has demonstrated improvements in track resolution yielding reduced gamma ray backgrounds~\cite{ferrario2019demonstration,kekic2021demonstration,simon2021boosting}.
\item NEXT has demonstrated viability of operation with diffusion-reducing gas mixtures~\cite{fernandes2020low,felkai2018helium,mcdonald2019electron}
\item NEXT has made significant progress on detecting a lone Ba ion within a Xe volume~\cite{mcdonald2018demonstration,NEXTBaTag2019,NEXTBaTag2021,rivilla2020fluorescent}.
\item PandaX has satisfactorily installed a module to evaluate the concept for \BBz~\cite{Pandax-4T-2022}.
\item PandaX has developed a radio-pure high pressure vessel~\cite{Wang_2020}.
\item PandaX should show a sub-1\% resolution. The Collaboration has reached 3\% FWHM~\cite{Wang_2020}.
\item \MJ\ has demonstrated robust radiopure cables and connectors~\cite{Arnquist2022MJDzero}.
\item \MJ\ demonstrated that its nearby parts are radiopure~\cite{Haufe2022,Gilliss2019}.
%
%
\item nEXO demonstrated a cathode voltage of 50~kV and an electric field of 400~V/cm~\cite{nEXO-pCDR}.
\item nEXO has identified two commercial photodetectors with high quantum efficiency for VUV photons~\cite{Adhikari_2021nEXO}.
\item nEXO has identified suitable high radiopurity, cryogenically compatible readout electronics~\cite{Adhikari_2021nEXO} and further improved the understanding of the anticipated background~\cite{EXO-200_bkgd}.
\item nEXO has made significant progress on detecting a lone Ba ion within a Xe volume~\cite{nEXOBaTag2019}.
\item nEXO has developed an electroformed copper vessel design, further reducing the projected backgrounds~\cite{Adhikari_2021nEXO}.

\item CUORE has shown efficient operation of a bolometric ton-scale detector~\cite{CUORE-0Te}.
\item CUPID has shown that the use of scintillating bolometers removes the $\alpha$ background~\cite{CUPID-MOzero}.
\item CUPID has shown that $\gamma$ background from internal parts of the detector is sufficiently low~\cite{CUPID-pCDR2019}.
\item CUPID has shown that cosmogenic background is sufficiently low~\cite{CUPID-pCDR2019}.
\item CUPID has shown that TeO$_2$ has an acceptable level of bulk radioactivity and the added light sensors do not add to the background. This was shown for the LMO crystals in Ref.~\cite{CUPID-pCDR2019}.
\item CUPID has found an acceptable supplier for their crystals. A few possible producers were identified in Ref.~\cite{CUPID-pCDR2019}.
\item 
Kamland-Zen has proposed a viable path to improving its energy resolution
by a factor of two with the KamLAND2-Zen upgrade~\cite{Shirai2017}.
\item Kamland-Zen has improved its \nuc{10}{C} tagging and reduced that background by an order of magnitude~\cite{Gando_2016,KamLAND-Zen2022}.
\end{enumerate}

\newpage

Table~\ref{tab:FutureExperiments} lists the large number of ongoing, proposed, or R\&D efforts in double beta decay. The variety of technologies and isotopes demonstrates the creativity and enthusiasm of this community.

\begin{table*}[!htbp]
\caption{A summary list of the \BBz\ ongoing and proposed experiments. The mass values are detector mass. The dark matter focused experiments (LZ, PandaX-4T, XENONnT, Darwin, R2D2) are optimized for those searches, but would have an opportunity to investigate \BBz, especially if they use enriched Xe. }
\begin{center}
\begin{tabular}{|c|c|c|c|c|c|}
\hline
\rowcolor{turquoiseblue}Experiment									&   Isotope		& Mass		&  Technique								& Present Status		& Location  \\
\hline
\hline
\rowcolor{LightCyan}CANDLES-III~\cite{CANDLES2021}					&  \nuc{48}{Ca}		& 305 kg		&$^{nat}$CaF$_2$ scint. crystals				& Operating 			&  Kamioka \\	
\rowcolor{SREblizzardblue}CDEX-1~\cite{Dai_2022}						&\nuc{76}{Ge}		& 	1 kg		&\nuc{enr}{Ge} semicond. det.					& Prototype			& CJPL		\\	
\rowcolor{SREblizzardblue}CDEX-300$\nu$~\cite{Dai_2022}				&\nuc{76}{Ge}		& 	225 kg	&\nuc{enr}{Ge} semicond. det.					& Construction			& CJPL		\\	
\rowcolor{SREblizzardblue}LEGEND-200~\cite{LEGEND-pCDR}			& \nuc{76}{Ge}		& 200 kg		&\nuc{enr}{Ge} semicond. det.					& Commissioning		& LNGS		\\ 
\rowcolor{SREblizzardblue}LEGEND-1000~\cite{LEGEND-pCDR}			& \nuc{76}{Ge}		& 1 ton		&\nuc{enr}{Ge} semicond. det.					& Proposal			&			\\ 
\rowcolor{LightCyan}CUPID-0~\cite{CUPID-0Se}						&\nuc{82}{Se}		& 10 kg		& Zn$^{enr}$Se scint. bolometers				& Prototype			& LNGS	\\	
\rowcolor{LightCyan}SuperNEMO-Dem~\cite{Arnold2021sNEMO}			&  \nuc{82}{Se}		& 7 kg		& \nuc{enr}{Se} foils/tracking					& Operation       	& Modane   \\ 
\rowcolor{LightCyan}SuperNEMO~\cite{Arnold2021sNEMO}				&  \nuc{82}{Se}		& 100 kg		& \nuc{enr}{Se} foils/tracking					& Proposal 		    	&  Modane   \\ 
\rowcolor{LightCyan}Selena~\cite{Selena2021}							&  \nuc{82}{Se}		& 			&\nuc{enr}{Se}, CMOS						& Development			& 		\\	
\rowcolor{LightCyan}IFC~\cite{JonesIFC2022}							&  \nuc{82}{Se}		& 			& ion drift SeF$_6$ TPC						& Development			& 		\\	
\rowcolor{SREblizzardblue}CUPID-Mo~\cite{CUPID-MOzero}				& \nuc{100}{Mo}	&	4~kg	& Li$^{enr}$MoO$_4$,scint. bolom.				& Prototype			&LNGS	\\	
\rowcolor{SREblizzardblue}AMoRE-I	~\cite{Lee_2020}					& \nuc{100}{Mo}	& 6 kg		& $^{40}$Ca$^{100}$MoO$_4$ bolometers		& Operation 			& YangYang  \\ 
\rowcolor{SREblizzardblue}AMoRE-II~\cite{Lee_2020}					& \nuc{100}{Mo}	& 200 kg		& $^{40}$Ca$^{100}$MoO$_4$ bolometers		& Construction			& Yemilab \\ 
\rowcolor{SREblizzardblue}CROSS~\cite{Armatol_2021}					& \nuc{100}{Mo}	&5~kg		& Li$_2$$^{100}$MoO$_4$, surf. coat bolom.		& Prototype			& Canfranc \\ 
\rowcolor{SREblizzardblue}BINGO~\cite{Armatol2022}					& \nuc{100}{Mo}	&			& Li$^{enr}$MoO$_4$						& Development			&LNGS	\\
\rowcolor{SREblizzardblue}CUPID~\cite{CUPID-pCDR2019}				& \nuc{100}{Mo}	&	450~kg	& Li$^{enr}$MoO$_4$,scint. bolom.				& Proposal			&LNGS	\\	
\rowcolor{LightCyan} China-Europe~\cite{Xue2019}						&  \nuc{116}{Cd}	& 			& \nuc{enr}{Cd}WO$_4$ scint. crystals			& Development    	         &   CJPL \\	
\rowcolor{LightCyan}COBRA-XDEM~\cite{Temminghoff2019}				&  \nuc{116}{Cd}	& 0.32 kg		& \nuc{nat}{Cd} CZT semicond. det.				& Operation            		 &  LNGS  \\	
\rowcolor{LightCyan}Nano-Tracking~\cite{BrownNanoTracking2019}			&  \nuc{116}{Cd}	& 			& \nuc{nat}{Cd}Te. det.						& Development            	 &    \\	
\rowcolor{SREblizzardblue}\textit{TIN.TIN}~\cite{Nanal2014}					& \nuc{124}{Sn}	&			& Tin bolometers							& Development			& INO	\\	
\rowcolor{LightCyan}CUORE~\cite{ADAMS2022CUORE}					&  \nuc{130}{Te}	& 1 ton		& TeO$_2$ bolometers						& Operating			&  LNGS            \\ 
\rowcolor{LightCyan}SNO+~\cite{Albanese_2021}						&  \nuc{130}{Te} 	& 3.9 t		& 0.5-3\% $^{nat}$Te loaded liq. scint.              		& Commissioning	   	&  SNOLab             \\ 
\rowcolor{SREblizzardblue}nEXO~\cite{Adhikari_2021nEXO}				&  \nuc{136}{Xe}	& 5 t			&Liq.  \nuc{enr}{Xe} TPC/scint.					& Proposal           		   &               \\	
\rowcolor{SREblizzardblue}NEXT-100~\cite{Romo_Luque_2022}			&  \nuc{136}{Xe} 	& 100 kg			& gas TPC                                        			& Construction	  		&   Canfranc            \\  
\rowcolor{SREblizzardblue}NEXT-HD~\cite{Romo_Luque_2022}			&  \nuc{136}{Xe} 	& 1 ton			& gas TPC                                        			& Proposal		 	&   Canfranc            \\	
\rowcolor{SREblizzardblue}AXEL~\cite{Ban2020AXEL}					&  \nuc{136}{Xe} 	& 				& gas TPC                                        			& Prototype	  		&   	            \\  
\rowcolor{SREblizzardblue}KamLAND-Zen-800~\cite{KamLAND-Zen2022}	& \nuc{136}{Xe} 	& 745 kg & \nuc{enr}{Xe} disolved in liq. scint.  					& Operating	   		& Kamioka \\ 
\rowcolor{SREblizzardblue}KamLAND2-Zen~\cite{Shirai2017}				& \nuc{136}{Xe}	& 		& \nuc{enr}{Xe} disolved in liq. scint.  				& Development	   		& Kamioka \\ 
\rowcolor{SREblizzardblue}LZ~\cite{LZzero2020}						& \nuc{136}{Xe}	&	600 kg	& Dual phase Xe TPC, nat./enr. Xe				& Operation			& SURF	\\	
\rowcolor{SREblizzardblue}PandaX-4T~\cite{Pandax-4T-2022}				& \nuc{136}{Xe}	& 3.7 ton		& Dual phase nat. Xe TPC					& Operation			& CJPL	\\	
\rowcolor{SREblizzardblue}XENONnT~\cite{Aprile_2022}					& \nuc{136}{Xe}	& 5.9 ton		& Dual phase Xe TPC 						& Operating			& LNGS	\\	
\rowcolor{SREblizzardblue}DARWIN~\cite{Agosstini2022DARWIN}			& \nuc{136}{Xe}	& 50 ton		& Dual phase Xe TPC 						& Proposal			& LNGS	\\	
\rowcolor{SREblizzardblue}R2D2~\cite{Bouet_2021}						& \nuc{136}{Xe}	& 			& Spherical Xe TPC 							& Development			& 	\\	
\rowcolor{SREblizzardblue}LAr TPC~\cite{MastbaumXeDopedLAr2022}		& \nuc{136}{Xe}	& kton		& Xe-doped LR TPC							&Development			&		\\
\rowcolor{LightCyan}NuDot~\cite{Graham_2019}						&	Various		&			& Cherenkov and scint. in liq. scint.				& Development			&		\\ 
\rowcolor{LightCyan}\textsc{Theia}~\cite{Askins_2020}						&	Xe or Te		&			& Cherenkov and scint. in liq. scint.				& Development			&		\\ 
\rowcolor{LightCyan}JUNO~\cite{JUNO2021}						&	Xe or Te		&			& Doped liq. scint.				& Development			&		\\ 
\rowcolor{LightCyan}Slow-Fluor~\cite{DungerSlowFluor2022}				&	Xe or Te		&			& Slow Fluor Scint.							& Development			&		\\ 
\hline
\end{tabular}
\end{center}
\label{tab:FutureExperiments}
\end{table*}

\clearpage
\bibliography{LRP2022BBWhitePaper}
\end{document}